\documentstyle[aps,aipbook,floats,epsf]{revtex}
\begin{document}


\def\beq{\begin{equation}}
\def\eeq{\end{equation}}
\def\bea{\begin{eqnarray}}
\def\eea{\end{eqnarray}}
\def\Csix{C^{(6)}}
\def\Ceight{C^{(8)}}
\def\costhstar{\cos\theta^*}
\def\Osix{O^{(6)}}
\def\Oeight{O^{(8)}}
\def\tbar{\bar t}
\def\shat{\hat s}
\def\that{\hat t}
\def\uhat{\hat u}
\def\CA{{\cal A}}
\def\GeV{{\>\, \rm GeV}}
\def\mtsq{m_t^2}
\def\third{{1\over3}}
\def\half{{1\over2}}
\def\mt{m_t}
\def\mperp{{m_\perp}}
\def\qbar{\bar q}
\def\pperp{p_\perp}
\def\Rang{R_{\rm ang}}
\def\Rrms{R_{\rm rms}}
\def\TeV{{\>\, \rm TeV}}

\title{Strong $W_L W_L$ Scattering$^\dagger$}

\author{R. Sekhar Chivukula$^*$}
\address{$^*$Department of Physics\\
Boston University \\
590 Commonwealth Avenue \\
Boston MA 02215 }

\address{$^\dagger$ BUHEP-95-17\\
hep-ph/9505202}

\maketitle

\begin{abstract}
I describe theories of a strongly-interacting electroweak symmetry
breaking sector and discuss the expected size of anomalous weak-boson
couplings in these models.
\end{abstract}

\section*{Signatures of Electroweak Symmetry Breaking in $WW$ Scattering}

The physics of electroweak symmetry breaking must appear at energies
of order a TeV or lower. To see this, consider a thought experiment
\cite{lqt}, the scattering of longitudinally polarized $W^+$ and
$W^-$:
\beq
{\lower20pt \hbox{\epsfxsize=3.5truein \epsfbox{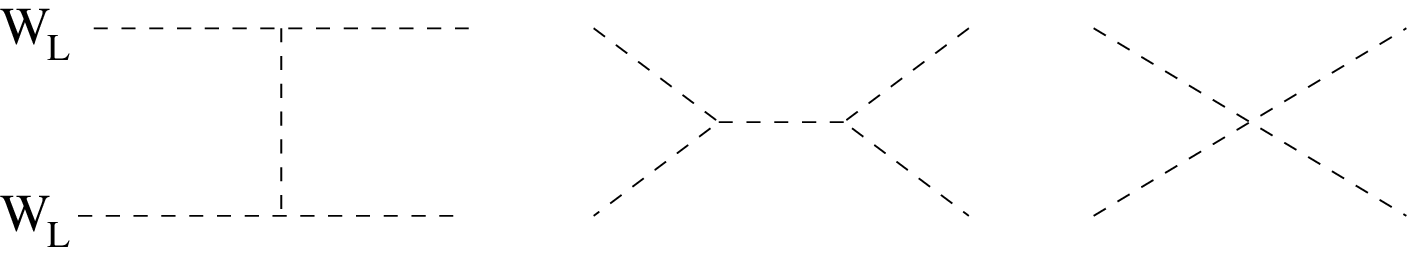}}}~~.
\eeq
Using the Feynman-rules of the electroweak gauge theory we can
calculate $W_L^+W_L^-$ scattering at tree level.  We find that this
amplitude grows like $E_{cm}^2$:
\beq
{\cal A} = {g^2 s \over 8 M_W^2}\left(1+\cos\theta^{*}\right)
{}~~~,
\eeq
plus terms that do not grow with $s$.
Projecting onto the $s$-wave state, we find
\beq
{\cal A}^{l=0} = {g^2 s \over 128\pi M_W^2}\sim
\left(\sqrt{s} \over 2.5\ {\rm TeV}\right)^2
{}~~~.
\eeq
Unitarity implies that the dynamics associated with EWSB has to appear
before an energy scale of around 2.5 TeV \cite{lqt} \cite{velt}.  There are
three
possibilities:
\begin{itemize}
\item There may be additional particles with masses less than or of
order of a TeV, or
\item the $W$ and $Z$ interactions may become strong at energies of order a
TeV, or
\item both of the above.
\end{itemize}

It is important to note that the amplitude calculated above {\it
universal} \cite{chan} \cite{golden}:
the calculation depended only on the {\it gauge structure} of
the standard model and on the relationship
\beq
\rho={M_W \over M_Z \cos\theta_W} \approx 1~~~,
\eeq
and will hold {\it regardless} of the dynamics responsible for
electroweak symmetry breaking. Therefore, in order to understand the
dynamics of electroweak symmetry breaking, it will be necessary to
characterize the physics which cuts-off the growth in the longitudinal
gauge boson scattering amplitudes.

The universality of the scattering amplitudes for
longitudinal gauge boson scattering can also be seen as a consequence of
the fact that the longitudinal components of the gauge-bosons are the
``eaten'' Goldstone Bosons of $SU(2) \times U(1)$ breaking. More
formally, the ``Equivalence Theorem''
\cite{equiv} \cite{chan} states that any amplitude
involving the scattering of longitudinal gauge-bosons is equal to the
same amplitude involving the corresponding Goldstone Bosons (which
would be present in the ungauged theory), up to corrections of order
$(M_W/E)^2$. The low-energy scattering amplitudes of Goldstone Bosons,
however, are determined by the low energy theorems of chiral dynamics
({\it c.f.} PCAC in QCD) and are determined by the symmetry structure
of the theory. In a theory in which $\rho=1$, the symmetry structure
of the electroweak symmetry breaking sector
is naturally $SU(2) \times SU(2)$
\cite{custodial}, and the scattering of the
longitudinal components of the $W$ and $Z$ are determined by
analogs of Weinberg's low-energy theorems in QCD \cite{golden}.

\section*{Theories of Electroweak Symmetry Breaking}

\subsection*{The Standard One-Doublet Higgs Model}

In the standard one-doublet Higgs model one introduces a fundamental
scalar doublet of $SU(2)_W$:
\beq
\phi=\left(\matrix{\phi^+ \cr \phi^0 \cr}\right)
{}~~~,
\eeq
which has a potential of the form
\beq
V(\phi)=\lambda \left(\phi^{\dagger}\phi - {v^2\over 2}\right)^2
{}~~~.
\eeq
In the potential, $v^2$ is assumed to be positive in order to
favor the generation of a non-zero vacuum expectation value for
$\phi$.  This vacuum expectation value breaks the electroweak
symmetry, giving mass to the $W$ and $Z$.  When symmetry breaking
takes place, the four degrees of freedom in $\phi$ divide up. Three of
them become the longitudinal components, $W_L$ and $Z_L$, of the gauge
bosons, and the fourth, commonly called $H$ (for Higgs particle), is left
over
\beq
\phi=\Omega\left(
\matrix{ 0\cr {{H+v}\over\sqrt{2}}\cr}
\right)
{}~~~.
\eeq
Here, $\Omega$ is an $SU(2)$ matrix.  If we make an $SU(2)_W$ gauge
transformation until $\Omega$ is the identity, we arrive at unitary gauge.

The exchange of the Higgs boson contributes to $W_L W_L$
scattering.  In the limit in which $E_{cm}$ is large compared to the
masses of the particles in the process, the leading contribution (in
energy) from Higgs boson exchange exactly cancels the bad high-energy
behavior in $W^+_L W^-_L$ scattering
\beq
{\lower15pt\hbox{\epsfxsize=2.2truein \epsfbox{fig2.eps}}}
\rightarrow
{\cal A} = -{g^2 s \over 8 M_W^2}\left(1+\cos\theta^{*}\right)
{}~~~,
\eeq
plus terms which do not grow with energy.

At tree-level the Higgs boson has a mass given by $m^2_H = 2\lambda
v^2$.  In order for this theory to give rise to strong $W$ and $Z$
interactions, it would be necessary that the Higgs boson be heavy and,
therefore, that $\lambda$ be large.

This explanation of electroweak symmetry breaking is unsatisfactory for a
number of reasons.  For one thing, this model does not give a dynamical
explanation of electroweak symmetry breaking. For another, when
embedded in theories with additional dynamics at higher energy scales,
these theories are technically unnatural \cite{thooft} .

Perhaps most unsatisfactory, however, is that theories of fundamental
scalars are probably ``trivial'' \cite{wilson}, {\it i.e.}, it is not possible
to construct an interacting
theory of scalars in four dimensions that is valid to arbitrarily
short distance scales.  In quantum field theories, fluctuations in the
vacuum screen charge -- the vacuum acts as a dielectric
medium. Therefore there is an effective coupling constant which
depends on the energy scale ($\mu$) at which it is measured. The
variation of the coupling with scale is summarized by the
$\beta$--function of the theory
\beq
\beta(\lambda) = \mu{d\lambda\over d\mu}
{}~~~.
\eeq
The only coupling in the Higgs sector of the standard model is the
Higgs self-coupling $\lambda$. In perturbation theory, the
$\beta$-function is calculated to be
\beq
{\lower15pt\hbox{\epsfysize=0.5 truein \epsfbox{fig4.eps}}}
\rightarrow \beta = {3\lambda^2 \over 2 \pi^2}
{}~~~.
\eeq
Using this $\beta$--function,
one can compute the behavior of the coupling constant as a function of
the scale\footnote{Since these expressions were computed in perturbation
theory, they are only valid when $\lambda(\mu)$ is sufficiently
small. We will return to the issue of strong coupling below.}. One
finds that the coupling at a scale $\mu$ is related to the coupling at
some higher scale $\Lambda$ by
\beq
{1\over\lambda(\mu)}={1\over\lambda(\Lambda)}
+{3\over 2\pi^2}\log{\Lambda\over\mu}
{}~~~.
\eeq
In order for the Higgs potential to be stable, $\lambda(\Lambda)$
has to be positive. This implies that
\beq
{1\over\lambda(\mu)} \ge {3\over
2\pi^2}\log{\Lambda\over\mu}
{}~~~.
\eeq
Thus, we have the bound
\beq
\lambda(\mu)\le{2\pi^2\over 3\log\left({\Lambda\over\mu}\right)}
{}~~~.
\eeq
If this theory is to make sense to arbitrarily short distances, and
hence arbitrarily high energies, we should take $\Lambda$ to $\infty$
while holding $\mu$ fixed at about 1 TeV. In this limit we see
that the bound on $\lambda$ goes to zero. In the continuum limit, this
theory is trivial; it is free field theory.

The inequality above can be translated into an upper bound on the mass
of the Higgs boson\cite{dn}. From the bound above, we have
\beq
{\Lambda\over\mu}\le \exp{\left(
{2\pi^2\over 3\lambda(\mu)}\right)}
{}~~~,
\eeq
but
\beq
m_H^2\sim 2v^2\lambda(m_H)
{}~~~,
\eeq
thus
\beq
\Lambda \le m_H \exp{\left({4\pi^2 v^2\over 3 m_H^2}\right)}
{}~~~.
\eeq
For a given Higgs boson mass, there is a {\it finite} cutoff energy at which
the description of the theory as a fundamental scalar doublet stops
making sense. This means that the standard one-doublet Higgs model can
only be regarded as an {\it effective} theory valid below this cutoff.

The theory of a relatively light weakly coupled Higgs boson, can be
self-consistent to a very high energy.  For example, if the theory is
to make sense up to a typical GUT scale energy, $10^{16}$ GeV, then
the Higgs boson mass has to be less than about 170 GeV \cite{maiani}.  In
this sense, although a theory with a light Higgs boson does not really
answer any of the interesting questions ({\it e.g.}, it does not
explain {\it why} $SU(2)_W\times U(1)_Y$ breaking occurs), the theory does
manage to postpone the issue up to higher energies.

The theory of a heavy Higgs boson ({\it i.e.} with a mass of about 1
TeV), however, does not really make sense.  Since we have computed the
$\beta$-function in perturbation theory, our answer is only reliable
at energy scales at which $\lambda(\mu)$ (as well as the Higgs boson
mass) is small. Fortunately, non-perturbative lattice calculations are
available. Early estimates \cite{kuti} indicated that if the theory was to
make sense up to 4 TeV, the mass of the Higgs boson had to be less
than about 640 GeV.  More recent results \cite{neuberger} imply that this
bound may be relaxed somewhat; one might be able to get away with an
800 GeV Higgs boson, but the Higgs boson mass is certainly bounded by
a value of this order of magnitude. The triviality limits on the mass
of the Higgs boson imply that it is not possible for the $W_L$ and
$Z_L$ scattering amplitudes in the standard model to truly become
large at energies well below the cutoff.  This result is especially
interesting because it implies that if nothing shows up below energies
of the order 700--800 GeV, then something truly ``non-trivial'' is
going on. We just have to find it.

\subsection*{Technicolor}

In models with fundamental scalars, electroweak symmetry breaking can
be accommodated if the parameters in the potential (which presumably
arise from additional physics at higher energies) are suitably
chosen. By contrast, technicolor theories strive to explain
electroweak symmetry breaking in terms of physics operating at an
energy scale of order a TeV.  In technicolor theories, electroweak
symmetry breaking is the result of chiral symmetry breaking in an
asymptotically-free, strongly-interacting gauge theory with massless
fermions.  Unlike theories with fundamental scalars, these theories
are technically natural: just as the scale $\Lambda_{QCD}$ arises
in QCD by dimensional transmutation, so too does the weak scale $v$ in
technicolor theories.  Accordingly, it can be
exponentially smaller than the GUT or Planck scales.  Furthermore,
asymptotically-free non-abelian gauge theories may be fully consistent
quantum field theories.

In the simplest technicolor theory \cite{weinberg} one introduces a
(massless) left-handed weak-doublet of ``technifermions'', and the
corresponding right-handed weak-singlets, which transform as $N$'s of
a strong $SU(N)_{TC}$ technicolor gauge group. In analogy to the
(approximate) chiral $SU(2)_L \times SU(2)_R$ symmetry on quarks in
QCD, the strong technicolor interactions respect an $SU(2)_L \times
SU(2)_R$ global chiral symmetry on the technifermions. When the
technicolor interactions become strong, the chiral symmetry is broken
to the diagonal subgroup, $SU(2)_{L+R}$, producing three
Nambu-Goldstone bosons which become, via the Higgs mechanism, the
longitudinal degrees of freedom of the $W_L$ and $Z_L$.  Because the
left-handed and right-handed techni-fermions carry different
electroweak quantum numbers, the electroweak interactions break to
electromagnetism.  If the $f$-constant of the theory, the analog of
$f_\pi$ in QCD, is chosen to be $v\approx 246$ GeV, then the $W$ mass
has its observed value. Furthermore \cite{custodial}, the remaining
$SU(2)_{L+R}$ custodial symmetry insures that, to lowest order in the
hypercharge coupling, $M_W = M_Z \cos\theta_W$.

In addition to the ``eaten'' Nambu-Goldstone bosons, such a theory
will give rise to various resonances, the analogs of the $\rho$,
$\omega$, and possibly the $\sigma$, in QCD.  In general, the growth
of the $W_L$ and $Z_L$ scattering amplitudes are cut off by
exchange of these heavy resonances,
\goodbreak\medskip
\vbox{
\epsfxsize=3.5in \epsfbox{fig5.eps}
\medskip\noindent
}
\medbreak
\noindent
just as in QCD the growth of pion--pion scattering amplitudes are cut
off by QCD resonances.  Scaling from QCD, we expect that the masses of
the various resonance will be of order a TeV.  Unlike the situation in
models with only fundamental scalars in the symmetry breaking sector,
the scattering of longitudinal $W$ and $Z$ bosons can truly be strong.

In figure 1, we show the data for the scattering amplitude of $\pi^+
\pi^0$ at low-energies, as well as the corresponding low-energy
theorem. We see that while the growth of the scattering amplitude
begins close to the low-energy theorem prediction, it is significantly
enhanced (and unitarized) by the presence of the $\rho$-resonance.
We expect a similar behavior in technicolor theories, with the energy
scale enhanced by a factor of $v/f_\pi \approx 2600$.

\begin{figure}[hb] 
\vspace*{-0.25in}
\epsfxsize=4.5in
\epsfbox[20 250 450 600]{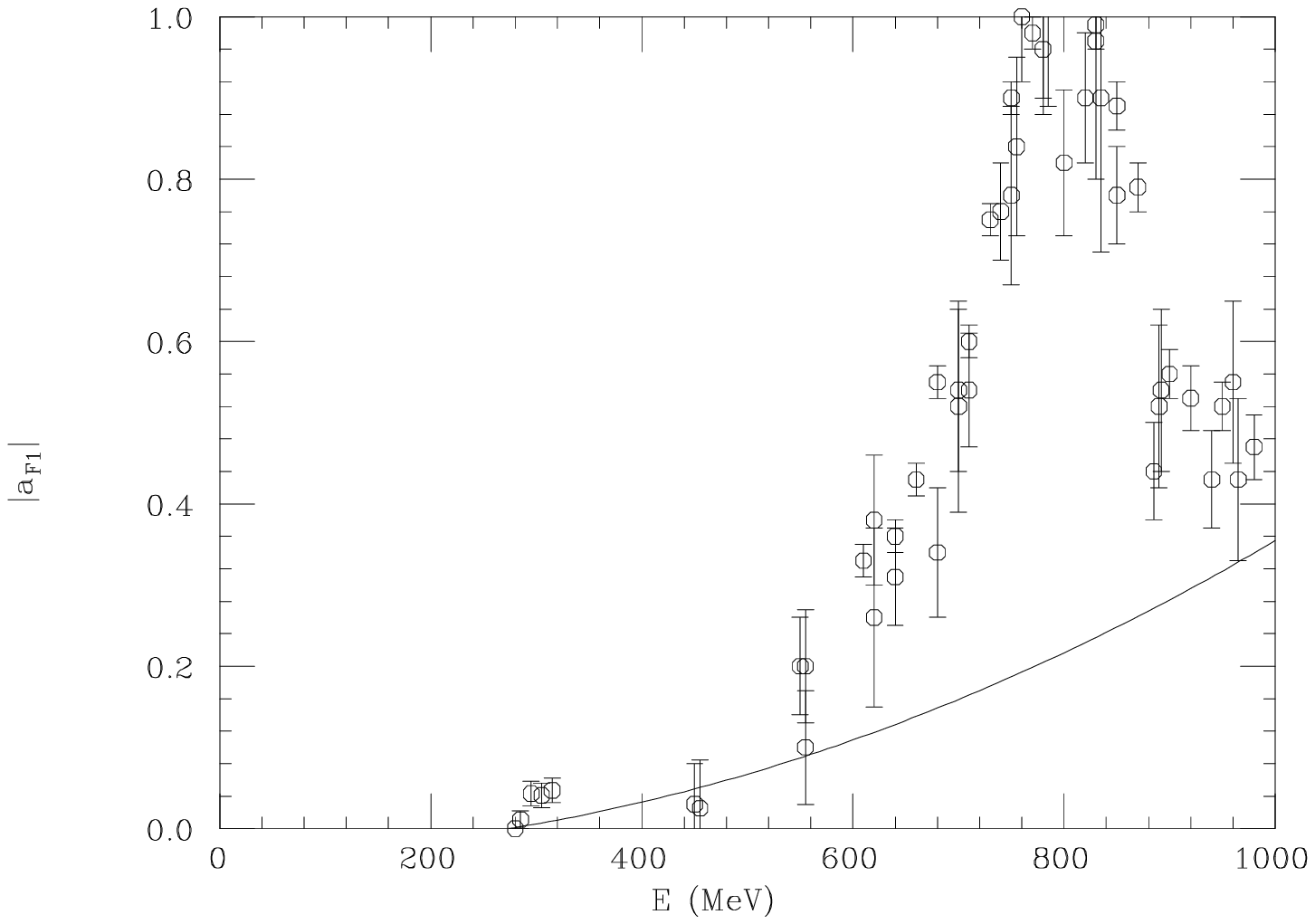}
\caption[figone]{Data \cite{dvr} and low-energy theorem
prediction for the spin-1/isospin-1 pion scattering amplitude.}
\end{figure}

The most direct signal for technicolor, therefore, is an enhancement
in the production of $WZ$ pairs at high invariant-mass, coming from
the production of the technicolor analog of the $\rho$-meson in QCD
\cite{ehlq} \cite{chan}. If the technirho resonance(s) are too heavy
to be observed at the LHC, there may be an enhancement in the
isospin-2 $W^+W^+ + W^- W^-$ channel which is large enough to be
observed \cite{berger} \cite{bagger}.  Detecting technicolor at the
LHC is likely to be quite challenging, however.  Recent estimates
\cite{kilgore} \cite{bagger} of the luminosity required to detect a
technicolor at the LHC indicate that it would be necessary to
accumulate of order 100 fb$^{-1}$, and that this would result in a
signal of only a few tens of events (over a background of comparable
size!).

\subsection*{Inelastic Channels in WW-Scattering}

Up to now, we have assumed that the {\it only} ``light'' particles in
the electroweak symmetry breaking sector are the longitudinal
components of the $W$ and $Z$. In a theory of this sort, the behaviors
described above are generic: the growth in the $W_L W_L$
scattering amplitude may be cut-off by light, narrow resonances (such
as in the weakly-coupled standard model) or by heavy, broad resonances
(such as would be expected in the simplest technicolor model).
However, if the global symmetry structure of the theory is larger than
$SU(2) \times SU(2)$, there may be additional (pseudo-)Goldstone
bosons. These additional particles give rise to {\it inelastic}
channels for vector-boson scattering, and may have dramatic
consequences for the behavior of the theory.

Consider a technicolor model with a global $SU(N_f)_L \times
SU(N_f)_R$ chiral symmetry which breaks spontaneously to the vectorial
$SU(N_f)$ subgroup, breaking the weak interactions and producing $N_f^2
- 1$ Goldstone (or pseudo-Goldstone) bosons.  The low energy theorem
for the $SU(N_f)$ singlet, spin singlet scattering amplitude of these
bosons is \cite{cahnsuz}
\beq
a_{singlet} = {N_f N_d s \over 32 \pi v^2},
\eeq
where $N_d$ is the number of technifermion doublets.  In analogy to
the analysis of the $W^+_L W^-_L$ scattering amplitude given at the
beginning of this talk, we see that as $N_f$ and $N_d$ increase, {\it
i.e.} as the number of inelastic channels in $W_L W_L$ scattering
grow, the scale by which the dynamics of EWSB must appear {\it
decreases} \cite{sundrum} \cite{chivgol}.  For example, in the one family
technicolor model \cite{onefamily} $N_f=8$ and $N_d=4$. In this model
$a_{singlet}$ would exceed unitarity at 440 GeV, and we expect that
new physics must appear at the energy scale or lower.

\begin{figure}[ht] 
\epsfxsize=4.5in
\epsfbox[90 230 480 525]{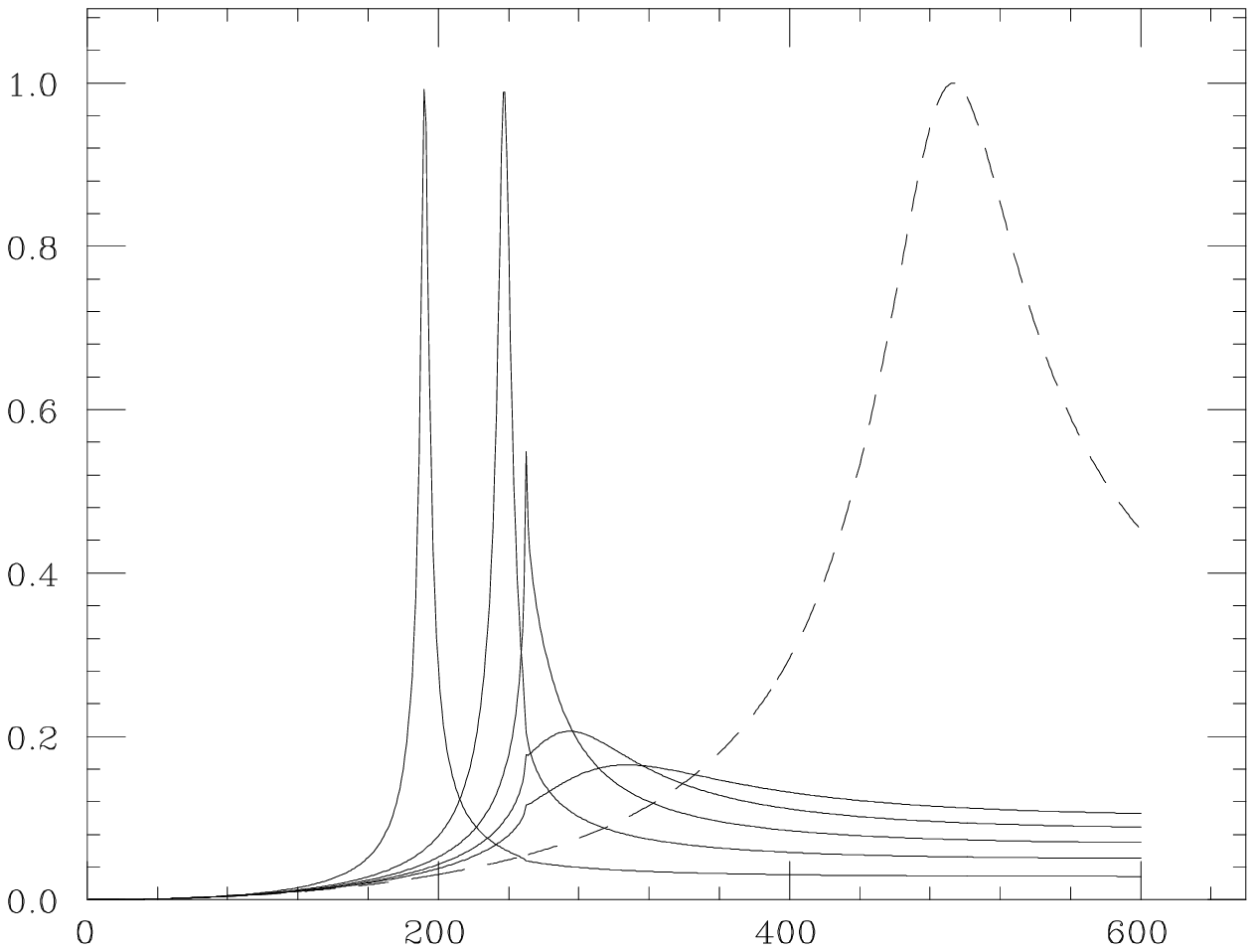}
\caption[figtwo]{The absolute value of the (weak) isospin-0 $W_L W_L$
scattering amplitude in a toy $O(N)$ model of
electroweak symmetry breaking \cite{chivgol}. The model contains 32
pseudo-Goldstone bosons, and the different solid curves show the
change in the amplitude as the mass of the pseudo goldstone bosons is
adjusted. The right-most nearly structureless amplitude corresponds to
the case where the ``Higgs'' in this model is strongly coupled, but
can decay to paris of pseudos in addition to pairs of weak gauge
bosons. The dashed-line corresponds to the same scattering amplitude
in the standard model with a 500 GeV Higgs boson.}
\end{figure}

Mitch Golden and I studied the phenomenology of a model of
electroweak-symmetry breaking with many inelastic channels in a
toy-model based on a scalar $O(N)$ theory \cite{chivgol}. We showed
that, although the new physics occurs at relatively low energies, this
new low-energy physics can be hard to detect. The presence of the
large numbers of inelastic channels can result in {\it elastic} $W$
and $Z$ scattering amplitudes that are small and structureless at all
energies, i.e. lacking any discernible resonances (see Fig. 2).
Nonetheless, the theory can be strongly interacting and the {\it
total} $W$ and $Z$ cross sections large: most of the cross section is
for the production of particles other than the $W$ or $Z$.  In such a
model, discovering the electroweak symmetry breaking sector will
depend on the observation of the other particles and our ability to
associate them with symmetry breaking.  This implies that we must keep
an open mind about the experimental signatures of the electroweak
symmetry breaking sector and that we cannot rely solely on two gauge
boson final states.

\section*{What Does This Imply for Anomalous Weak-Boson Self-Interactions?}

It would seem natural that in a theory of a strongly interacting
electroweak symmetry breaking sector, like technicolor, there could be
large corrections to the electroweak gauge-boson self-couplings. For
example, one would expect that the coupling of one longitudinal
gauge-boson to two transverse gauge-bosons would acquire a form-factor
similar to the electromagnetic form-factor of the pion in QCD.
\beq
{\lower20pt \hbox{\epsfxsize=2.0truein \epsfbox{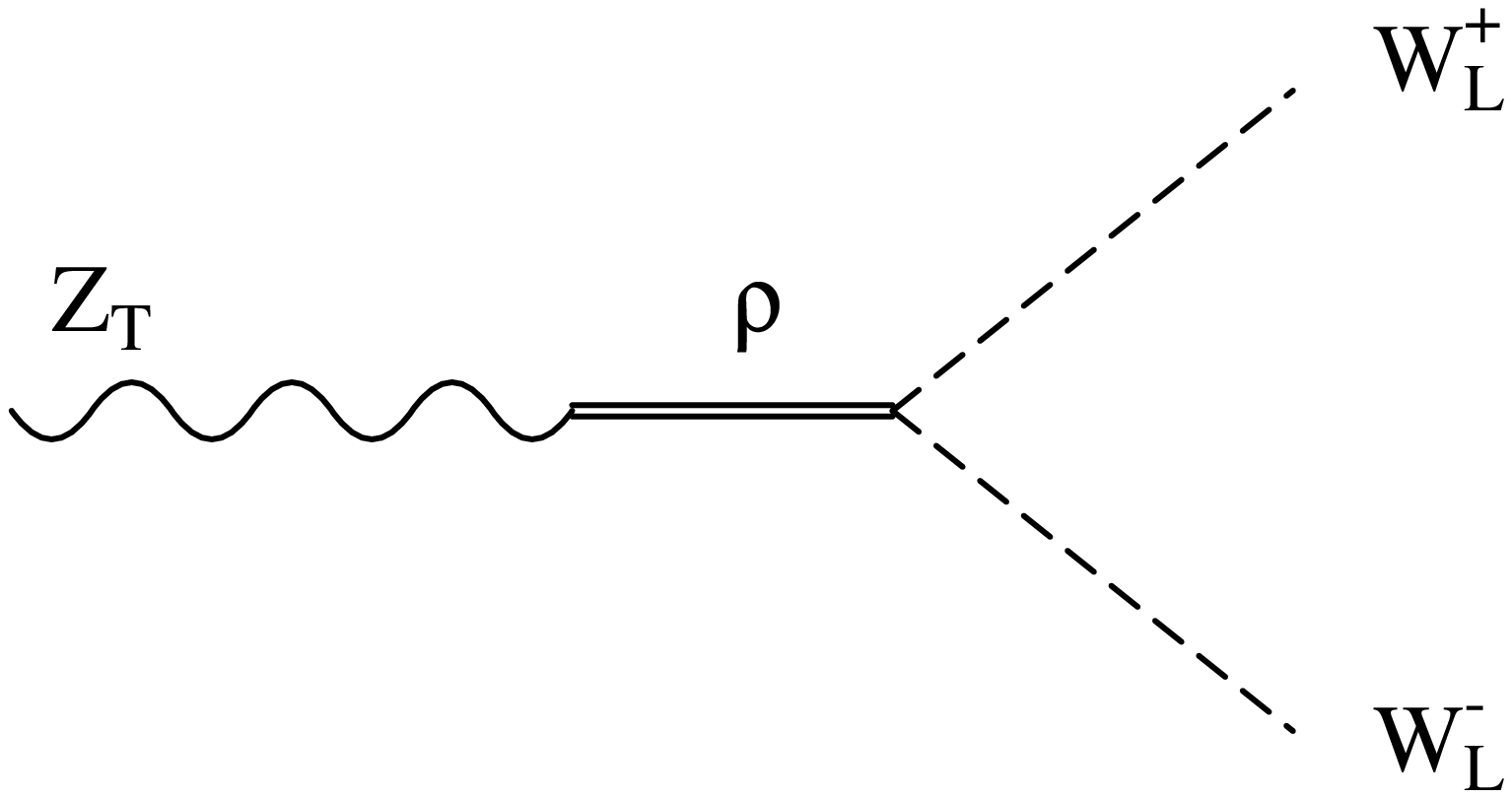}}}
\eeq

As discussed by Wudka at this conference \cite{wudka}, one can use
dimensional analysis to estimate the size of the corrections to
the weak-boson self-couplings \cite{hagiwara}
\beq
\Delta g_1,~ \Delta\kappa = {\cal O}\left( {g^2  \over 16\pi^2}\right)~~,
\eeq
and
\beq
\lambda = {\cal O}\left( {g^4 \over (16\pi^2)^2}\right)~~,
\eeq
Using these estimates, we see that deviations in $\kappa$ and $g_1$ are
expected to be of order $10^{-3}$, while $\lambda$ is
expected to be of order $10^{-5}$ or $10^{-6}$.

I would like to emphasize here that these dimensional estimates have a
simple physical interpretation in terms of the form-factor picture
that I discussed above (see also the discussion of Willenbrock at this
conference \cite{willenbrock}). As in QCD, we expect that the scale of
variation of the form factor is given by the mass of the lowest-lying
resonance in in the appropriate channel, namely by the mass of a
vector meson.  Furthermore, in the limit that $M_W,\ M_Z \to 0$, we
know that the vector bosons must couple to a conserved current and
that the vector-boson self couplings must be of canonical form
\cite{weinbgauge}.  Therefore, we can estimate that the size of the
anomalous couplings $\kappa$ and $g$ must be
\beq
\Delta g_1, ~ \Delta\kappa = {\cal O}\left( {M^2_W
\over M^2_{\rho_{TC}}}\right)
\eeq
and, given that $\lambda$ is the coefficient of a {\it dimension-6}
operator and the normalization chosen in \cite{hagiwara},
\beq
\lambda = {\cal O}\left( {M^4_W \over M^4_{\rho_{TC}}}\right)~.
\eeq
Afficianados of dimensional analysis \cite{NDA} will see immediately
that these two estimates are, in fact, consistent since the
dimensional analysis estimate of the lightest-resonance mass in models
of electroweak symmetry breaking are of order $4\pi v$.

What are the prospects for the experimental detection of deviations of
this size?  Baur, Han, and Ohnemus \cite{baur} have recently
considered this issue for a variety of colliders. The prospects
are discouraging. For example, for the LHC with an integrated
luminosity of 100 fb$^{-1}$, they find that one may be able to probe
to the level of 10$^{-2}$ for $\Delta g$ and $\lambda$. This is
not sufficient to be sure to probe effects of the size predicted above.

On the other hand, one might wonder if the analysis of the effects of
anomalous weak-boson self-interactions is consistent with the results
given in the previous section. From the analysis given above, we see
that a $\Delta g$ to order 10$^{-2}$ would arise in a model with a
technirho of mass approximately 1 TeV. This is consistent with the
analysis of \cite{kilgore}: in both cases one is looking for the
effects of technirho mesons on the production of $WZ$ pairs!

\section*{conclusions}

A strongly interacting symmetry breaking sector will result in one or
more resonances which are either:
\begin{itemize}

\item Heavy (with masses of order a TeV) and broad (in the case that
elastic $W$ and $Z$ scattering dominates). Detection will require an
integrated luminosity of order 100 fb$^{-1}$ or more at the LHC.

\item Light and broad (in the case that inelastic channels are
important).  In this case detection will hinge on observing particles
other than the $W_L$ and $Z_L$ and identifying them as being
associated with EWSB.

\end{itemize}

In the first case, the lightest resonances in the electroweak symmetry
breaking sector are expected to be the technivector mesons, the
analogs of the $\rho$ and $\omega$ in QCD. The masses of these
resonances are expected to be of order a TeV, and one expects an
enhancement of $WZ$ and/or $WW$ production at energies of this order
of magnitude. One may think of the  ``tail'' of the technirho as giving
rise to anomalous weak-boson self-interactions.  The expect size of
the resulting anomalous gauge boson vertices is small, with
$\Delta\kappa$ and $\Delta g$ of order $10^{-3}$ and $\lambda$ of
order $10^{-5}$ or $10^{-6}$.

\section*{Acknowledgements}
I gratefully acknowledge the support of NSF Presidential Young
Investigator Award and a DOE Outstanding Junior Investigator Award.

{\it This work was supported in part by the National Science
Foundation under grant PHY-9057173
and  by the Department of Energy under
contract DE-FG02-91ER40676.}

\vfill\eject

\end{document}